\documentclass[preprint,showpacs,showkeys,pra]{revtex4}
\usepackage{epsf} \usepackage{color}
\usepackage{amsmath,amsthm,amssymb}
\usepackage{dcolumn}
\usepackage{bm}
\everymath{\displaystyle} \numberwithin{equation}{section}
\begin{document} 

\title{Foldy-Wouthyusen wave functions and conditions of
transformation between Dirac and Foldy-Wouthuysen
representations}

\author{V. P. Neznamov}
\email{neznamov@vniief.ru} \affiliation{Russian Federal Nuclear
Center -- The All-Russian Research Institute of Experimental
Physics, Sarov 607190, Russia}

\author{A. J. Silenko}
\email{silenko@inp.minsk.by} \affiliation{Research Institute of
Nuclear Problems, Belarusian State University, Minsk 220030,
Belarus}

\date{\today}

\begin {abstract}
The
block-diagonalization of the Hamiltonian is not sufficient
for the transformation to the Foldy-Wouthuysen (FW) representation. The
conditions enabling the transition from the Dirac representation to the
FW one are formulated and proved. The connection between wave
functions in the two representations is derived. The results obtained allows
calculating expectation values of operators
corresponding to main classical quantities.
\end{abstract}

\pacs {03.65.-w, 11.10.Ef, 12.20.-m} \keywords{relativistic
quantum mechanics; Foldy-Wouthyusen transformation; unitary
transformations} \maketitle

\section{Introduction}

The Foldy-Wouthuysen representation introduced in Ref. \cite{FW}
occupies a special place in the quantum theory. This is mainly due
to the fact that the FW representation provides the best
possibility of obtaining a meaningful classical limit of the
relativistic quantum mechanics \cite{FW,CMcK,JMP,PRA}. The
Hamiltonian and all operators in this representation are
block-diagonal (diagonal in two spinors). The basic advantages of
the FW representation are investigated in Refs.
\cite{FW,CMcK,JMP,PRA} and shortly described in Sec. \ref{Sec2}.

There are many methods of the FW transformation considered in
Refs. \cite{JMP,PRA,E,VJ,VANT,FizElem} (see also below). However,
some of them are rather intuitive. Therefore, conditions of the FW
transformation and main properties of the FW representation should
be determined.

In the present work, the basic properties of wave functions in the
FW representation are investigated and the connection between wave
functions in the Dirac and FW representations is found. Such a
connection has been determined in Ref. \cite{PEPANFW} in the
particular case when the FW transformation is exact.

We also establish conditions of the transformation from the Dirac
representation to the FW one. The action of these conditions is
illustrated by several examples. In particular, the obtained
result allows calculating the expectation values of operators
corresponding to the basic classical quantities. Explicit forms of
these operators for relativistic particles in external fields can
be determined in the FW representation but not in the Dirac one.

\section{Foldy-Wouthuysen representation} \label{Sec2}

The use of the FW representation possesses the important
advantages investigated in Refs. \cite{FW,CMcK,JMP}. The relations
between the operators in the FW representation are similar to
those between the respective classical quantities. Only the FW
representation possesses these properties considerably simplifying
the transition to the semiclassical description. The Hamiltonian
for a free particle fully agrees with that of classical physics in
contrast with the Hamiltonian in the Dirac representation
\cite{FW}.

We use the system of units $\hbar =c=1$ and denote matrices as
follows. $$\begin{array}{c} \bm{\alpha}=\beta\bm\gamma=
\left(\begin{array}{cc} 0  &  \bm{\sigma} \\ \bm{\sigma} & 0
\end{array}\right), ~~ {\beta}\equiv\gamma^0=\left(\begin{array}{cc} 1  &  0
\\ 0 & -1 \end{array}\right), ~~  \bm{\Sigma}
=\left(\begin{array}{cc} \bm{\sigma}  &  0 \\ 0 &
\bm{\sigma}\end{array}\right),   ~~\bm{\Pi}=\beta\bm\Sigma
=\left(\begin{array}{cc} \bm{\sigma}  &  0 \\ 0 &
-\bm{\sigma}\end{array}\right)  \end{array}  $$ are the Dirac
matrices; $\sigma ^i$ are the Pauli matrices. $\psi _D (x),~\psi
_{FW} (x)$ are four-component wave functions in the Dirac and FW
representation, respectively. The
scalar product of
four-vectors is taken in the form $xy\equiv x^\mu y_\mu \equiv
x^0y^0-x^ky^k,~\mu =0,1,2,3,~k=1,2,3;~p^\mu =i(\partial /\partial
x_\mu ).$

The position operator in the FW representation is $\bm {r}$. It is
given by the cumbersome expression in the Dirac representation
\cite{NW}:
$$ \bm r_D=\bm r+\frac{i\beta\bm\alpha}{2E}-
\frac{i\beta(\bm\alpha\cdot\bm p)\bm p+[\bm\Sigma\times\bm p]|\bm
p|} {2E(E+m)|\bm p|}, ~~~E=\sqrt{m^2+\bm p^2}.
$$ For free particles, the momentum and velocity operators are
expressed in a normal, close to classical, way, $\bm p=-i\nabla$
and $\bm v=\bm p/E$ ($\bm v=\bm \alpha$ in the Dirac
representation).

In the FW representation, the problem of "\textit{Zitterbewegung}"
motion never arises \cite{CMcK,NW}. The operators $\bm l=\bm
r\times\bm p$ and $\bm\Sigma/2$ define the angular momentum and
the spin for a free particle, respectively. In this
representation, unlike the Dirac one, each of them is a constant
of motion (see Ref. \cite{FW}). The corresponding operators
conserving in the Dirac representation are known only for free
particles and are expressed by cumbersome formulae (see Refs.
\cite{FW,JMP}). The FW representation is very convenient for
describing the particle polarization. In this representation,
polarization operators have simple forms. The three-dimensional
polarization operator is equal to the matrix $\bm\Pi$
\cite{FG,TKR}. In the Dirac representation, this operator depends
on the particle momentum \cite{JMP,FG,TKR}. For particles
interacting with external fields, it also depends on external
field parameters \cite{TKR}.

Thus, in the Dirac representation all operators corresponding to
the basic classical quantities are defined by cumbersome
expressions. Moreover, these operators should also depend on the
external field parameters for particles interacting with external
fields. Therefore, explicit
forms of such operators in the Dirac representation are usually
unknown. We can conclude that the preferable employment of the FW
representation is evident, although the relativistic wave
equations are more complicated in this representation.

The use of the FW representation permits to reduce the number of components
of the bispinor wave function because one of the FW spinors is zero.

Note also that the derivation of the relativistic Hamiltonian
${\cal H}_{FW}$ in Ref. \cite{VANT} made it possible to treat
quantum-field processes in the FW representation within the
framework of the perturbation theory \cite{FizElem}.

Equations for the wave function in the FW representation have a
non-covariant form, and the FW Hamiltonians are non-local and
block-diagonal (they contain infinite sets of differential
operators and are diagonal in two spinors).

\section{Methods of the Foldy-Wouthuysen transformation}

In this Section, we give an overview of known methods of the FW
transformation.

In the presence of time-dependent external fields, transformation
to the FW representation described by the wave function
$\psi_{FW}$ is performed with the unitary operator $U_{FW}$:
$$\psi_{FW}=U_{FW}\psi_{D}=e^{iS}\psi_{D},$$
where $\psi_{D}=A\left(\begin{array}{c} \phi \\ \chi
\end{array}\right)$ is the wave function (bispinor) in the Dirac
representation.

The Hamilton operator in the FW representation takes the form
\cite{FW,Gol}:
\begin{equation} {\cal H}_{FW}=U_{FW}{\cal H}_{D}U_{FW}^{-1}-iU_{FW}\frac{
\partial U_{FW}^{-1}}{\partial t}. \label{eq3.1} \end{equation}

The Dirac Hamiltonian can be split into operators commuting and
noncommuting with the operator $\beta$:
\begin{equation} {\cal H}_{D}=\beta m+{\cal E}+{\cal
O},~~~\beta{\cal E}={\cal E}\beta, ~~~\beta{\cal O}=-{\cal
O}\beta. \label{eq3.2} \end{equation} The Hamiltonian ${\cal
H}_{D}$ is Hermitian. We assume that both operators ${\cal E}$ and
${\cal O}$ are also Hermitian.

In the classical work by Foldy and Wouthuysen \cite{FW}, the exact
transformation for free relativistic particles and the approximate
transformation for nonrelativistic particles in electromagnetic
fields have been carried out.

For free Dirac particles, ${\cal E}=0,~{\cal
O}=\bm{\alpha}\cdot\bm{ p}$.

The FW Hamiltonian $({\cal H}_0)_{FW}$ and the FW wave function
$\psi_{FW}(x)$ are related to the Dirac Hamiltonian for free
particles,
\begin{equation} ({\cal H}_0)_{D}=\beta m+\bm{\alpha}\cdot\bm{ p},
\label{eq3.3} \end{equation} and to the Dirac wave function
$\psi_{D}(x)$ by the unitary transformation
\begin{equation} ({\cal H}_0)_{FW}=U_{0}({\cal H}_0)_{D}U_{0}^{\dag}=\beta E,
~~~\psi_{FW}(x)=U_{0}\psi_{D}(x),
~~~U_{0}=\sqrt{\frac{E+m}{2E}}\left(1+\frac{\beta\bm{\alpha}\cdot\bm{
p}}{E+m}\right). \label{eq3.4} \end{equation}

The FW equation for the free motion of a quantum mechanical
particle with the spin 1/2 takes the following form:
\begin{equation} p_0\psi_{FW}(x)=({\cal H}_0)_{FW}\psi_{FW}(x)
=\beta E\psi_{FW}(x). \label{eq3.5} \end{equation}

Solutions of Eq. (\ref{eq3.5}) are plane waves with positive and
negative energy \cite{FW}:
\begin{equation} \psi_{FW}^{(+)}(x)=\frac{1}{(2\pi)^{3/2}}{\cal
U}e^{-ipx},~~~\psi_{FW}^{(-)}(x)=\frac{1}{(2\pi)^{3/2}}{\cal
V}e^{ipx},~~~p_0=E=\sqrt{m^2+\bm p^2}. \label{eq3VI}
\end{equation}
In Eq. (\ref{eq3VI}), ${\cal U}=\left(\begin{array}{c} \phi \\
0
\end{array}\right)$ and ${\cal V}=\left(\begin{array}{c} 0 \\
\chi
\end{array}\right)$, $\phi$ and $\chi$ are two-component normalized
Pauli spin functions.

For ${\cal U}$ and ${\cal V}$, the following orthonormality and
completeness conditions are true:
\begin{equation}\begin{array}{c} {\cal U}_s^{\,\dag} {\cal U}_{s'}={\cal V}_s^{\,\dag} {\cal V}_{s'}=
\delta_{ss'}, ~~~ {\cal U}_s^{\,\dag} {\cal V}_{s'}={\cal
V}_s^{\,\dag}
{\cal U}_{s'}=0,\\
\sum\limits_s{({\cal U}_s)_\gamma({\cal
U}_{s}^{\,\dag})_\delta}=\frac12(1+\beta)_{\gamma\delta}, ~~~
\sum\limits_s{({\cal V}_s)_\gamma({\cal
V}_{s}^{\,\dag})_\delta}=\frac12(1-\beta)_{\gamma\delta}.
\end{array} \label{eq3.7}
\end{equation}
In expressions (\ref{eq3VI}), (\ref{eq3.7}), $\gamma,\delta$
belong to spinorial indices and $s$ to spin indices. Further, the
sum sign and the indices themselves are not shown, when summing is
performed in the spinorial indices.

It has been shown in Ref. \cite{JMP} that the exact FW
transformation can be fulfilled with the operator
\begin{equation}
U_{FW}=\sqrt{\frac{E+m}{2E}}\left(1+\frac{\beta {\cal
O}}{E+m}\right),~~~ E=\sqrt{m^2+{\cal O}^2} \label{eq3.8}
\end{equation} on condition that the operators ${\cal E}$ and
${\cal O}$ commute:
\begin{equation}
[{\cal E},{\cal O}]=0. \label{eq3.9}
\end{equation}
In this case, the FW Hamiltonian is given by
\begin{equation}
{\cal H}_{FW}=\beta E+{\cal E}. \label{eq3.10}
\end{equation}
Eq. (\ref{eq3.10}) agrees with formula (\ref{eq3.5}) for free
particles.

Condition (\ref{eq3.9}) is satisfied for the Dirac equation at a
presence of the static external magnetic field $\bm
B=\operatorname{rot}\bm A$ (exact transformation by Case \cite{C})
and also of some other interactions described in Refs.
\cite{JMP,RMZ}. In the general case of
interaction of a fermion with an
arbitrary boson field, condition (\ref{eq3.9}) is not satisfied
and the problem of transition between the Dirac and FW
representations becomes much more complicated.

The general form of the exact FW transformation has been found by
Eriksen \cite{E} for stationary external fields. The Eriksen
transformation operator is given by \cite{E}
\begin{equation}
U_{E}=U_{FW}=\frac12(1+\beta\lambda)\left[1+\frac14(\beta\lambda+\lambda\beta-2)\right]^{-1/2},
~~~ \lambda=\frac{{\cal H}_D}{({\cal H}_D^2)^{1/2}}, \label{eq3XI}
\end{equation}
where ${\cal H}_{D}$ is the Hamiltonian in the Dirac
representation. $\lambda=+1$ and
$-1$ for the positive-energy and negative-energy solutions,
respectively. It is important that \cite{E}
\begin{equation}\lambda^2=1, ~~~ [\beta\lambda,\lambda\beta]=0 \label{eq3.12}
\end{equation}
and the operator $\beta\lambda+\lambda\beta$ is even:
\begin{equation}[\beta,(\beta\lambda+\lambda\beta)]=0.\label{eq3XIII}
\end{equation}

Any even operator is block-diagonal and does not mix the upper and
lower components of the wave function.

The validity of the Eriksen transformation has 
been argued in Ref. \cite{VJ}. It has been shown that the Eriksen
transformation directly leads to the FW representation. We can
give an additional argument which follows from the fact that the
use of Eqs. (\ref{eq3.12}),(\ref{eq3XIII}) reduces transformation
operator (\ref{eq3XI}) to the form
\begin{equation}
U_{E}=(2+\beta\lambda+\lambda\beta)^{-1/2}(1+\beta\lambda).
\label{eq3.14}
\end{equation}
Since two factors in the right hand side of Eq. (\ref{eq3.14})
commutate and the first factor defines an even operator, an action
of $U_{E}$ on any eigenfunction of the Dirac Hamiltonian nullifies
either the lower spinor or the upper one. Eq. (\ref{eq3.14}) can
also be transformed to the form
\begin{equation}
U_{E}=\frac{1+\beta\lambda}{\sqrt{(1+\beta\lambda)^\dag(1+\beta\lambda)}}.
\label{eq3.15}
\end{equation}

The Eriksen operator brings the Dirac wave function and the Dirac
Hamiltonian to the FW representation in one step. However, it is
difficult to use the Eriksen method because the general final
formula is very cumbersome and contains roots of Dirac matrix
operators.

Another direct way to obtain the FW transformation has been
proposed in Ref. \cite{VANT} (see also overview \cite{FizElem}).
The transformation operator $U_{FW}$ and the relativistic
Hamiltonian ${\cal H}_{FW}$ (\ref{eq3.1}), obtained for the
general case with arbitrary external boson field as a power series
in coupling constant $q$, have the following form:
\begin{equation}\begin{array}{c}
U_{FW}=U_0(1+q\delta_1+q^2\delta_2+q^3\delta_3+\dots),\\
{\cal H}_{FW}=\beta E+qK_1+q^2K_2+q^3K_3+\dots
\end{array} \label{eq3.16}
\end{equation}
In expressions (\ref{eq3.16}), $U_0$ is the FW transformation
operator for free Dirac particles defined by Eq. (\ref{eq3.4}) and
$\delta_i,K_i$ are some operators. The Hamiltonian ${\cal H}_{FW}$
(\ref{eq3.16}) can be used, in particular, to consider field
quantum theory issues \cite{FizElem}.

Along with direct derivation of the block-diagonal Hamiltonians,
there is a lot of step-by-step methods to obtain Hamiltonians free
of odd operators. In particular, one of such methods has been used
in the classical work by Foldy and Wouthuysen \cite{FW} to
derive the Hamilatonian ${\cal
H}_{FW}$ in the presence of an external electromagnetic field as a
power series in $1/m$. In the next section, we compare
step-by-step and direct methods of transition to the FW
representation.

\section{Comparison of direct and step-by-step methods of the Foldy-Wouthyusen transformation}

The transformation of the Hamiltonian to a block-diagonal form may
not be equivalent to the FW transformation. There are infinitely
many representations that differ from the FW representation
despite the block-diagonal form of the Hamiltonian \cite{PRD}. As an example, one can indicate the
Eriksen-Kolsrud method \cite{EK} which variants have been used in
Refs. \cite{Ob1,Ob2,Heidenreich,Ob3}. It has been proved in Ref.
\cite{PRD} (see also below) that these transformations
are not equivalent to the FW
transformation. The same conclusion will be made for the Melosh
transformation \cite{Me}.

One should draw special attention to the step-by-step method
initially proposed in the classical work by Foldy and Wouthuysen
\cite{FW} and widely used in many applications. De Vries and
Jonker \cite{VJ} and later the author of Ref. \cite{VANT} have
shown that the step-by-step removal of odd operators being the
main distinguishing feature of step-by-step methods does not
result in the FW representation. The Hamiltonians transformed to
the FW representation by the Eriksen method \cite{E} and the
method developed in Ref. \cite{VANT} differ from the Hamiltonian
obtained by the original FW method \cite{FW}. A reason is a
noncommutativity of unitary transformations \cite{VANT,FizElem}.
Formula (\ref{eq3.1}) can be re-written as
$$
{\cal H}_{FW}=U_{FW}\left({\cal H}_{D}-i\frac{\partial}{\partial
t}\right)U_{FW}^{-1}+i\frac{\partial}{\partial t}.
$$
Any unitary transformation operator can be presented in the
exponential form
$$
U_{FW}=e^{iS}.
$$
For direct methods \cite{E,VANT}
\begin{equation}
{\cal H}_{FW}=e^{iS}\left({\cal H}_{D}-i\frac{\partial}{\partial
t}\right)e^{-iS}+i\frac{\partial}{\partial t}. \label{eq4.1}
\end{equation}
For step-by-step methods
\begin{equation}
{\cal H}_{FW}=\dots e^{iS_n}\dots e^{iS_2}e^{iS_1}\left({\cal
H}_{D}-i\frac{\partial}{\partial t}\right)e^{-iS_1}e^{-iS_2}\dots
e^{-iS_n}\dots +i\frac{\partial}{\partial t}. \label{eq4.2}
\end{equation}
Hamiltonians (\ref{eq4.1}), (\ref{eq4.2}) are equivalent only when
\begin{equation}
e^{iS}=e^{i(S_1+S_2+\dots S_n+\dots)}=\dots e^{iS_n}\dots
e^{iS_2}e^{iS_1}. \label{eq4.3}
\end{equation}
However, equality (\ref{eq4.3}) is valid only for the trivial case
of commuting $S_i$'s. Such a situation almost never takes place in
applications.

According to the theorem of F. Haussdorff \cite{Haussdorff},
\begin{equation}
\exp{A}\cdot\exp{B}=\exp{\left(A+B+\frac12[A,B]+higher ~order~
commutators\right)}\neq\exp{B}\cdot\exp{A}. \label{eq4.4}
\end{equation} The noncommutativity of unitary transformations
leads to a dependence of the resulting operator of the FW
transformation
\begin{equation}
U=U_{FW}=\dots U_n\dots U_2U_1U_0 \label{eq4.5}
\end{equation}
on a specific method of this transformation \cite{VANT,FizElem}.
This circumstance does not mean that the exact FW representation
cannot be reached in several steps. If the step-by-step
transformation has been carried out with operator (\ref{eq4.5}),
the Hamiltonian obtained can be brought to the exact FW form with
the unitary operator $U'=U_{E}U^{-1}$, where $U_{E}$ is given by
Eq. (\ref{eq3XI}). Evidently, the
exact FW representation needs one of the exact methods even in
this case.

It has been shown in Refs. \cite{VANT,VJ} that the original
step-by-step transformation \cite{FW} does not exactly lead to the
FW representation. The same situation takes place for the methods
developed in Ref. \cite{JMP,PRA}. Thus, step-by-step methods are
approximate and the exact FW
representation cannot be obtained with these methods.

However, step-by-step methods are rather helpful, when one can
restrict oneself to several leading orders of a FW Hamiltonian
expansion in a chosen small parameter. As a rule, this level of
accuracy is quite sufficient, in particular, when one uses the
weak field approximation or the nonrelativistic one. Eqs.
(\ref{eq4.2}),(\ref{eq4.4}) show that a difference between
Hamiltonians obtained by the exact and step-by-step methods is
defined by the commutator $[S_1,S_2]$. Therefore, this difference
appears for the first time only for the third step.

For example, the expansion in powers of $1/m$ in the
nonrelativistic approximation carried out in Ref. \cite{FW} gives
an accuracy up to $1/m^2$ and can be restricted to $S_1$ and
$S_2$. In this case, $e^{iS_2}e^{iS_1}\approx e^{i(S_1+S_2)}$,
since $[S_1,S_2]\sim 1/m^3$.

An essential advantage of the step-by-step methods consists in the relative
simplicity of computations they offer. On the contrary, the use of direct methods leads to cumbersome derivations.

The differences between the FW Hamiltonians obtained by the direct
and step-by-step methods are beyond the weak field approximation
and even beyond the leading terms of expansion in the Planck
constant. The latter statement is illustrated by the
correspondence between the quantum-mechanical motion equations
obtained by the method proposed in Ref. \cite{PRA} and respective
classical relativistic equations.

In addition to the aforesaid, any transformation that diagonalizes
the Hamiltonian and claims to enable the transition to the FW
representation should be tested for the wave function reduction
condition. The formulation and proof of this condition is provided
in the next section.

\section{Connection between wave functions in the Dirac and Foldy-Wouthyusen representations.
Proof of the wave function reduction condition for the Foldy-Wouthyusen transformation}

Diagonalization of the Hamiltonian relative to the upper and lower
components of the wave function $\psi_D(x)$ is the necessary
condition for the transformation from the Dirac representation to
the FW representation (the FW transformation).

The second condition for the FW transformation consists in the
nullification of either upper or lower components of the bispinor
wave function $$\psi_{D}(\bm x,t)=A\left(\begin{array}{c} \phi(\bm x,t) \\
\chi(\bm x,t)
\end{array}\right)$$ and the transformation
of the normalization operator of the wave function $\psi_D(x)$
into the unit operator. Let us call this the ``wave function
reduction condition''. For the case when the Dirac Hamiltonian is
independent of time (the free case or the case of static external fields)
this condition can be represented in the following form
\begin{equation}\begin{array}{c}
\psi_{D}^{(+)}(\bm x,t)=e^{-iEt}A_+\left(\begin{array}{c} \phi^{(+)}(\bm x) \\
\chi^{(+)}(\bm x)\end{array}\right)\rightarrow \psi_{FW}^{(+)}(\bm
x,t)=e^{-iEt}
\left(\begin{array}{c} \phi^{(+)}(\bm x) \\
0\end{array}\right);\\
\psi_{D}^{(-)}(\bm x,t)=e^{iEt}A_-\left(\begin{array}{c} \phi^{(-)}(\bm x) \\
\chi^{(-)}(\bm x)\end{array}\right)\rightarrow \psi_{FW}^{(-)}(\bm
x,t)=e^{iEt}
\left(\begin{array}{c} 0 \\
\chi^{(-)}(\bm x)\end{array}\right).
\end{array}
\label{eq5.1}
\end{equation}
In this equation, $E$ is the
module of the particle energy operator; $A_+$ and $A_-$ are
normalization operators, which may differ, in
general, for solutions with positive and negative energies.
Definition of the operators $A_+$
and $A_-$ implies that the wave functions $\psi_{D}^{(+)}(\bm
x,t),\psi_{D}^{(-)}(\bm x,t)$ and the spinors $\phi^{(+)}(\bm
x),\chi^{(-)}(\bm x)$ are normalized to 1:
$$\int{{\psi_{D}^{(\pm)}}^\dag(\bm
x,t)\psi_{D}^{(\pm)}(\bm x,t)dV}=1, ~~~ \int{{\phi^{(+)}}^\dag(\bm
x)\phi^{(+)}(\bm x)dV}=1, ~~~ \int{{\chi^{(-)}}^\dag(\bm
x)\chi^{(-)}(\bm x)dV}=1.$$ Pluses and minuses denote positive and
negative energy states, respectively.

For a free particle,
\begin{equation}
E=\sqrt{m^2+\bm p^2}, ~~~
A_+=A_-=\sqrt{\frac{E+m}{2E}}.\label{eq5.2}
\end{equation}
$\phi^{(+)}(\bm x)=e^{i\bm
p\cdot\bm x}\phi$ and $\chi^{(-)}(\bm x)=e^{-i\bm p\cdot\bm
x}\chi$ for the positive and negative energy solutions,
respectively. $\phi$ and $\chi$ are the two-component Pauli spin
functions [see Eq. (\ref{eq3VI})].

Functions $\psi_{D}^{(\pm)}(\bm x,t)$ and
$\psi_{FW}^{(\pm)}(\bm x,t)$ are the appropriate solutions
of the initial Dirac equation and the equation transformed to the FW
representation for a free particle and a particle moving in static external
fields. The reduction condition implies transformation of the Dirac
wave functions to the form
$\psi_{FW}^{(\pm)}(\bm x,t)$ with the unit normalization operator.

In general, the Dirac and FW Hamiltonians depend on time. In this
case, the reduction condition (\ref{eq5.1}) has the same meaning.
We use expansions in the Dirac equation solutions obtained either
for free motion of particles, or for
a motion in the presence of
static external fields when solving specific problems of physics
(at least, with the use of the perturbation theory).

The reduction condition can be proved with
the Eriksen
transformation \cite{E}, which is the exact FW transformation of
the time-independent Hamiltonian ${\cal H}_{D}$.
Let us consider the positive energy solutions.
Since $\lambda\psi_{D}^{(\pm)}(\bm
x,t)=\pm\psi_{D}^{(\pm)}(\bm x,t)$, Eriksen transformation
operator (\ref{eq3.14}) transforms
the Dirac wave function to the
form
\begin{equation}
\psi_{FW}^{(+)}(\bm
x,t)=e^{-iEt}\left[\frac12+\frac14\left(\beta\lambda+\lambda\beta\right)\right]^{-1/2}A_+
\left(\begin{array}{c} \phi^{(+)}(\bm x)
\\ 0
\end{array}\right).
\label{eq5.8}
\end{equation}
The wave function normalization requirement can be written as
\begin{equation}
\int{{\psi_{FW}^{(+)}}^\dag(\bm x,t)\psi_{FW}^{(+)}(\bm
x,t)}dV=\int{{\phi^{(+)}}^\dag(\bm
x)A_+\left[\frac12+\frac14\left(\beta\lambda+\lambda\beta\right)\right]^{-1}A_+\phi^{(+)}(\bm
x)}dV=1. \label{eq5.9}
\end{equation}
Eq. (\ref{eq5.9}) is valid only when the following condition is
satisfied:
\begin{equation}
A_+\left[\frac12+\frac14\left(\beta\lambda+\lambda\beta\right)\right]^{-1}A_+=1.
\label{eq5.10}
\end{equation}
Multiplying the left-hand and right-hand sides of equation
(\ref{eq5.10}) by the operator $A_+^{-1}$ and extracting a root
square results in
$$
\left[\frac12+\frac14\left(\beta\lambda+\lambda\beta\right)\right]^{-1/2}=A_+^{-1}
$$
and
\begin{equation}
A_+=\left[\frac12+\frac14\left(\beta\lambda+\lambda\beta\right)\right]^{1/2}.
\label{eq5XI}
\end{equation}
Eqs. (\ref{eq5.8}),(\ref{eq5XI}) prove the reduction condition
(\ref{eq5.1}).
An explicit form of the operator $A_+$ has been determined in Ref.
\cite{PEPANFW} in the particular case when the FW transformation
is exact.

Similar derivation proves Eq. (\ref{eq5.1}) for the negative
energy solutions. In this case
\begin{equation}
A_-=A_+=\left[\frac12+\frac14\left(\beta\lambda+\lambda\beta\right)\right]^{1/2}.
\label{eq5.12}
\end{equation}

\section{Verification of methods of the Foldy-Wouthuysen
transformation}

The reduction condition can be successfully used for the
verification of methods of the Foldy-Wouthuysen transformation.
The validity of the Eriksen method \cite{E} has been proved in the
precedent section. Next subsection is devoted to the method
proposed in Ref. \cite{VANT} and
discussed in Ref. \cite{FizElem}.

\subsection{Particle in a static electric field}

For a particle in a static electric field, we can demonstrate that
the condition (\ref{eq5.1}) is satisfied up to linear terms on $e$
and quadratic terms on $v/c$ in the expansion of $U_{FW}$ in
powers of charge $e$ \cite{VANT}. This procedure can, apparently,
be applied up to any order of expansion on $e$ and $v/c$ using the
mathematical technique of Ref. \cite{VANT}.

We obtain with denotations used in \cite{VANT} and within the
accepted accuracy that
\begin{equation}\begin{array}{c}
{\cal H}_D=\beta m+\bm \alpha\cdot\bm p+eA_0(\bm x), \\
U_{FW}=(1+\delta_1^0+\delta_1^e)U_0=1+\frac{\beta\bm\alpha\cdot\bm
p}{2m}-\frac{p^2}{8m^2}-\frac{ie}{4m^2}(\bm\alpha\cdot\nabla
A_0)
\\
-\frac{ie\beta}{16m^3}\left[(\bm \alpha\cdot\bm
p)(\bm\alpha\cdot\nabla A_0)-(\bm\alpha\cdot\nabla A_0)(\bm
\alpha\cdot\bm p)\right],\\ {\cal H}_{FW}=\beta E,
~~~E=m+\frac{p^2}{2m}+e\beta\left\{A_0+\frac{i}{8m^2}\left[(\bm
\alpha\cdot\bm p)(\bm\alpha\cdot\nabla A_0)-(\bm\alpha\cdot\nabla
A_0)(\bm \alpha\cdot\bm p)\right]\right\},\\
\psi_D^{(+)}(\bm
x,t)=e^{-iEt}\Bigl\{1-\frac{p^2}{8m^2}-\frac{ie}{16m^3}\left[(\bm
\sigma\cdot\bm p)(\bm\sigma\cdot\nabla A_0) \right.\\
\left. -(\bm\sigma\cdot\nabla A_0)(\bm \sigma\cdot\bm
p)\right]\Bigr\}\left(\begin{array}{c} \phi^{(+)}(\bm x)
\\ \left(\frac{\bm\sigma\cdot\bm
p}{2m}+\frac{ie\bm\sigma\cdot\nabla
A_0}{4m^2}\right)\phi^{(+)}(\bm x)
\end{array}\right),\\\psi_{FW}^{(+)}(\bm
x,t)=U_{FW}\psi_D^{(+)}(\bm x,t)=e^{-iEt}\left(\begin{array}{c}
\phi^{(+)}(\bm x)
\\ 0
\end{array}\right),\\
\psi_D^{(-)}(\bm
x,t)=e^{iEt}\Bigl\{1-\frac{p^2}{8m^2}+\frac{ie}{16m^3}\left[(\bm
\sigma\cdot\bm p)(\bm\sigma\cdot\nabla A_0) \right.\\
\left. -(\bm\sigma\cdot\nabla A_0)(\bm \sigma\cdot\bm
p)\right]\Bigr\}\left(\begin{array}{c}
-\left(\frac{\bm\sigma\cdot\bm p}{2m}
-\frac{ie\bm\sigma\cdot\nabla A_0}{4m^2}\right)\chi^{(-)}(\bm x)
\\ \chi^{(-)}(\bm x)
\end{array}\right),\\\psi_{FW}^{(-)}(\bm
x,t)=U_{FW}\psi_D^{(-)}(\bm x,t)=e^{iEt}\left(\begin{array}{c}
0
\\ \chi^{(-)}(\bm x)
\end{array}\right).
\end{array}
\label{eqVI1}
\end{equation}

In Eq. (\ref{eqVI1}), $\bm p$ and functions of $\bm p$ imply the
corresponding operators; $\phi^{(+)}(\bm x)$ and $\chi^{(-)}(\bm
x)$ are two-component spinors.

It follows from (\ref{eqVI1}) that reduction condition
(\ref{eq5.1}) is satisfied and, therefore, the obtained unitary
transformation is the FW one.

The method described in Refs. \cite{VANT,FizElem} can also be
checked by means of its comparing with the Eriksen method
\cite{E}. In the case considered
(see Ref. \cite{Dirac})
$$\begin{array}{c}
\lambda=\frac{{\cal H}_D}{\sqrt{{\cal H}_D^2}}= \frac{\beta m+\bm
\alpha\cdot\bm p}{p_0}-\frac12{\cal B}+\frac{1}{2p_0}(\beta m+\bm
\alpha\cdot\bm p){\cal B}(\beta m+\bm \alpha\cdot\bm
p)\frac{1}{p_0},\\p_0=\sqrt{m^2+\bm p^2},~~~ p_0{\cal B}+{\cal
B}p_0=-2eA_0.
\end{array}$$
Since $${\cal
B}=-\frac{eA_0}{m}+\frac{e(p_0^2A_0+A_0p_0^2)}{2m^3}$$ to within
terms of order of $p^2/m^{2}$,
then
$$\begin{array}{c}\lambda=\beta +\frac{\bm \alpha\cdot\bm
p}{m}-\beta\frac{p^2}{2m^2}-\frac{i\beta e}{2m^2}\bm
\alpha\cdot\nabla A_0+ \frac{ie}{4m^3}\left[(\bm \alpha\cdot\nabla
A_0)(\bm \alpha\cdot\bm p)-(\bm \alpha\cdot\bm p)(\bm
\alpha\cdot\nabla A_0)\right],\\
\left(\frac{\beta
\lambda+\lambda\beta}{4}+\frac{1}{2}\right)^{-1/2}=1+\frac{
p^2}{8m^2}+\frac{i\beta e}{16m^3}\left[(\bm \alpha\cdot\bm p)(\bm
\alpha\cdot\nabla A_0)-(\bm \alpha\cdot\nabla A_0)(\bm
\alpha\cdot\bm p)\right],\\
U_E=1+\frac{\beta \bm \alpha\cdot\bm p}{2m}-\frac{
p^2}{8m^2}-\frac{ie}{4m^2}\bm \alpha\cdot\nabla A_0-\frac{i\beta
e}{16m^3}\left[(\bm \alpha\cdot\bm p)(\bm \alpha\cdot\nabla
A_0)-(\bm \alpha\cdot\nabla A_0)(\bm \alpha\cdot\bm
p)\right].\end{array}$$

The latter expression for $U_E$ coincides with the expression
(\ref{eqVI1}) for the FW transformation operator $U_{FW}$ obtained
by the method described in Refs. \cite{VANT,FizElem}.

\subsection{Super-algebra in the Dirac equation with static external
fields}

In Ref. \cite{RMZ}, supersymmetric quantum mechanics was applied
to a wide range of interactions between a Dirac particle and
external static fields that provides a closed form of a
block-diagonal Hamiltonian. The authors of Ref. \cite{RMZ}
considered the $SU(2)$ transformation of the Dirac Hamiltonian as
the FW transformation.

It is interesting to determine whether reduction condition
(\ref{eq5.1}) is satisfied for this transformation. Using
denotations from \cite{RMZ}, we have
\begin{equation}
{\cal H}_D=Q+Q^\dag+\Lambda, \label{eqfirst}
\end{equation}
where $\Lambda$ is a Hermitian operator, $Q$ and $Q^\dag$ are two
fermion operators satisfying the following requirements:
\begin{equation}
Q^2={Q^\dag}^2=0,~~~\{Q,\Lambda\}=\{Q^\dag,\Lambda\}=0.\label{eqsecond}
\end{equation}
$\{\dots,\dots\}$ denotes an
anticommutator.

Then, the Hermitian operators of $SU(2)$ algebra are introduced as
follows:
$$J_1=\frac{Q+Q^\dag}{2(\{Q,Q^\dag\})^{1/2}}, ~~~
J_2=\frac{-i\Lambda(Q+Q^\dag)}{2(\Lambda^2\{Q,Q^\dag\})^{1/2}},
~~~J_3=\frac{\Lambda}{2(\Lambda^2)^{1/2}},
~~~[J_i,J_j]=ie_{ijk}J_k.$$

The transformation operator is
\begin{equation}
U_{FW}=e^{iJ_2\theta}=\cos{\frac{\theta}{2}}+2iJ_2\sin{\frac{\theta}{2}}=
\sqrt{\frac12\left(1+\cos{\theta}\right)}+\frac{\Lambda(Q+Q^\dag)}{(\Lambda^2\{Q,Q^\dag\})^{1/2}}
\sqrt{\frac12\left(1-\cos{\theta}\right)}. \label{eqVI2}
\end{equation}
In contrast to Ref. \cite{RMZ}, the exponent of $iJ_2\theta$ in
Eq. (\ref{eqVI2}) is taken with the positive sign. This is a
necessary step to be in agreement with the reduction condition
(\ref{eq5.1}) (see below). Since
\begin{equation}
{\cal H}_{FW}=e^{iJ_2\theta}{\cal
H}_De^{-iJ_2\theta}=(Q+Q^\dag)\cos{\theta}+2iJ_2\Lambda\sin{\theta}+
\Lambda\cos{\theta}+2iJ_2(Q+Q^\dag)\sin{\theta} \label{eqVI3}
\end{equation}
and $$
\sin{\theta}=\frac{\{Q,Q^\dag\}^{1/2}}{(\{Q,Q^\dag\}+\Lambda^2)^{1/2}},~~~
\cos{\theta}=\frac{(\Lambda^2)^{1/2}}{(\{Q,Q^\dag\}+\Lambda^2)^{1/2}},~~~
\tan{\theta}=\frac{\{Q,Q^\dag\}^{1/2}}{(\Lambda^2)^{1/2}},$$
expression (\ref{eqVI3}) is reduced to the diagonal form
\begin{equation}
{\cal
H}_{FW}=\frac{\Lambda}{(\Lambda^2)^{1/2}}\left(\{Q,Q^\dag\}+\Lambda^2\right)^{1/2}.
\label{eqVI4}
\end{equation}
If \begin{equation}\Lambda=\beta m, ~~~ Q=\left(\begin{array}{cc} 0 &
 0\\ M & 0 \end{array}\right), ~~~ Q^\dag=\left(\begin{array}{cc} 0 &
 M^\dag \\ 0 & 0 \end{array}\right) \label{definition}
\end{equation}
and $$M=\bm\sigma\cdot(\bm p+\bm C)-iC_5, ~~~
C_i=A_i-i\varepsilon_i, ~~~ i=1,2,3,5,$$
 the Dirac Hamiltonian ${\cal H}_D$ is given by
\begin{equation}
{\cal H}_D=\beta m+\bm\alpha\cdot\bm\pi+i\beta\gamma_5\pi_5,
~~~\pi_i=p_i+A_i(\bm x)+i\beta\varepsilon_i(\bm
x),~~~i=1,2,3,5,~~~p_5=0. \label{eqVI5}
\end{equation}
The following interactions are described using these denotations:
i) $A_5$ is the pseudo-scalar potential, ii) $\varepsilon_5$ is
the time component of the axial-vector potential, iii)
$\bm\varepsilon$ is the ``electrical'' component of interaction of
the anomalous magnetic moment of
the particle, iv) $\bm A$ is the
minimum magnetic interaction; v) if $A_5=0,~\varepsilon_5=0,~\bm
A=0,~\bm\varepsilon=\bm r$, the Hamiltonian ${\cal H}_D$ reduces
to the Hamiltonian of the Dirac oscillator. All of the above
interactions admit a closed transformation to the diagonal form
(\ref{eqVI4}).

Let us check whether the $SU(2)$ transformation (\ref{eqVI2})
satisfies the reduction condition
(\ref{eq5.1}) and is the FW transformation.
In the case defined by Eq. (\ref{definition}),
\begin{equation}\begin{array}{c}
{\cal H}_{FW}=\beta E=\beta(\{Q,Q^\dag\}+m^2)^{1/2},
~~~E^2=\{Q,Q^\dag\}+m^2=\left(\begin{array}{cc} M^\dag
M+m^2 \!&\! 0 \\ 0 \!&\! MM^\dag+m^2
\end{array}\right), \\
U_{FW}=\sqrt{\frac{E+m}{2E}}\left[1+\frac{\beta(Q+Q^\dag)}{E+m}\right],\\
\psi^{(+)}_{D}(\bm
x,t)=e^{-iEt}\sqrt{\frac{E+m}{2E}}\left(\begin{array}{c}
\phi^{(+)}(\bm x)
\\ \frac{1}{E+m}M\phi^{(+)}(\bm x)
\end{array}\right), ~~~\psi^{(+)}_{FW}(\bm
x,t)=e^{-iEt}\left(\begin{array}{c} \phi^{(+)}(\bm x)
\\ 0\end{array}\right), \\
\psi^{(-)}_{D}(\bm
x,t)=e^{iEt}\sqrt{\frac{E+m}{2E}}\left(\begin{array}{c}
-\frac{1}{E+m}M^\dag\chi^{(-)}(\bm x)
\\ \chi^{(-)}(\bm x)
\end{array}\right), ~~~\psi^{(-)}_{FW}(\bm
x,t)=e^{iEt}\left(\begin{array}{c} 0
\\ \chi^{(-)}(\bm x) \end{array}\right).
\end{array}\label{eqVI.VI}
\end{equation}
Expressions (\ref{eqVI.VI}) show that, indeed, the reduction
condition (\ref{eq5.1}) is fulfilled for $SU(2)$ transformation
\cite{RMZ} (with the changed sign in the exponential factor
$iJ_2\theta$). Thus, this is a FW transformation. If the authors'
sign in this factor \cite{RMZ} remains unchanged
($e^{-iJ_2\theta}$), the transformation operator $U_{FW}$ in Eq.
(\ref{eqVI.VI}) takes the form
$$U=\sqrt{\frac{E+m}{2E}}\left[1-\frac{\beta(Q+Q^\dag)}{E+m}\right].$$
In this case, the reduction condition (\ref{eq5.1}) is violated
despite the block-diagonalization of the Hamiltonian.

\subsection{Eriksen-Kolsrud transformation}

The Eriksen-Kolsrud (EK) transformation \cite{EK} was used in many
works (see, e.g., Refs. \cite{Ob1,Ob2,Heidenreich,Ob3,N}). It is
fulfilled in two stages defined by the operators $U_{1}$ and
$U_{2}$. The unitary operator of resulting transformation is given
by \cite{EK}
\begin{equation}\begin{array}{c}
U_{EK}=U_{1}U_{2},~~~
U_{1}=\frac{1}{\sqrt2}\left(1+J\lambda\right),~~~
U_{2}=\frac{1}{\sqrt2}\left(1+\beta J\right),\\ J=i\gamma_5\beta,
~~~\lambda=\frac{{\cal H}_D}{({\cal H}_D^2)^{1/2}}.
\end{array} \label{eqVIVII}
\end{equation}

There are many examples of the exact EK transformation. It is
often claimed that this transformation is equivalent to the FW
one. It has been proved in Ref. \cite{PRD} that this statement is
incorrect. We can show that the EK transformation does not satisfy
reduction condition (\ref{eq5.1}).

For a free particle, ${\cal H}_{EK}=\beta E$ and
\begin{equation}\begin{array}{c}
\psi^{(+)}_{EK}(\bm x,t)=U_{EK}\psi^{(+)}_{D}(\bm
x,t)=e^{-iEt}\sqrt{\frac{E+m}{2E}}\left(\begin{array}{c}
\left(1+\frac{i\bm\sigma\cdot\bm p}{E+m}\right)\phi^{(+)}(\bm x)
\\ 0\end{array}\right), \\
\psi^{(-)}_{EK}(\bm
x,t)=e^{iEt}\sqrt{\frac{E+m}{2E}}\left(\begin{array}{c} 0
\\ \left(1-\frac{i\bm\sigma\cdot\bm p}{E+m}\right)\chi^{(-)}(\bm x)
\end{array}\right).\end{array}\label{eqVI8}
\end{equation}
It can be seen from Eq. (\ref{eqVI8}) that Eq. (\ref{eq5.1}) is
not satisfied and one needs to perform additional transformation
\cite{PRD}
\begin{equation}
U_{EK\rightarrow FW}=\sqrt{\frac{E
+m}{2E}}\left(1-\frac{i\beta\bm\sigma\cdot\bm p}{E+m}\right)
\label{eqVI9}
\end{equation}
which does not change the form of the Hamiltonian (${\cal
H}_{EK}={\cal H}_{FW}$).

Let us consider a Dirac particle in an external gravitational
field defined by the static metric $ds^2=V^2(\bm
x)(dx^0)^2-W^2(\bm x)d\bm x^2$. This problem was investigated in
Refs. \cite{PRD,Ob1,Ob2,Heidenreich}. In the considered case, the
EK transformation is exact \cite{Ob1,Ob2}.

The Dirac Hamiltonian is given by \cite{Ob1,Ob2}
\begin{equation}{\cal H}_{D}=\beta mV+\frac12\{\bm\alpha\cdot\bm p,{\cal
F}\},~~~ {\cal F}=\frac VW. \label{eqVI.10}
\end{equation}
Similarly to Refs. \cite{PRD,Ob1,Ob2}, we take into account the
first-order terms for the potentials $(V-1)$, $({\cal F}-1)$ and
their first-order spatial derivatives. We use the method proposed
in Refs. \cite{VANT,FizElem}. We perform the expansion in a power
series in $|\bm p|/m$ and take into account
terms up to $p^2/m^2$.

In this case
$$\begin{array}{c}
U_{FW}=1+\frac{\beta\bm\alpha\cdot\bm p}{2m}-\frac{p^2}{8m^2}+
\frac{\beta}{4m}\left\{({\cal F}-V),\bm\alpha\cdot\bm p\right\}
\\
-\frac{1}{16m^2}\left[({\cal F}-V)p^2+2\bm\alpha\cdot\bm p\,({\cal
F}-V)\,\bm\alpha\cdot\bm p+p^2({\cal F}-V)\right]
\end{array}$$
and
\begin{equation}\begin{array}{c}
{\cal H}_{FW}=\beta m+\frac{\beta p^2}{2m}+\beta m(V-1)-
\frac{\beta}{4m}\left\{p^2,(V-1)\right\}+\frac{\beta}{2m}\left\{p^2,({\cal
F}-1)\right\}\\-
\frac{\beta}{8m}\left[2\bm{\Sigma}\cdot(\bm\phi\times\bm
p)+\nabla\cdot\bm\phi\right]+\frac{\beta}{4m}\left[2\bm{\Sigma}\cdot(\bm
f\times\bm p)+\nabla\cdot\bm f\right], \end{array}\label{eqVI.XI}
\end{equation}
where $\bm\phi=\nabla V,~\bm f=\nabla{\cal F}$. Eq.
(\ref{eqVI.XI}) coincides with the corresponding equation obtained
in Ref. \cite{PRD}.

It can be shown that the FW wave functions satisfy Eq.
(\ref{eq5.1}). The EK transformation operator can be written as
\begin{equation}\begin{array}{c}
U_{EK}=\frac12\left\{1+i\gamma_5-\frac{\beta\gamma_5}{2}\{\bm\Sigma\cdot\bm
p,{\cal
F}\}\frac{1}{mV}-\frac{i\gamma_5}{4m^2}\left(\frac1Wp^2{\cal
F}+{\cal F}p^2\frac1W\right)\frac1V \right.\\
\left. -\frac{i\gamma_5}{4m^2}\left[2\bm{\Sigma}\cdot(\bm
f\times\bm p)+\nabla\cdot\bm
f\right]-\frac{i\gamma_5\beta}{2m}\bm{\Sigma}\cdot\bm\phi
-\frac{\gamma_5}{4m^2}\{\bm\Sigma\cdot\bm p,{\cal
F}\}\bm{\Sigma}\cdot\bm\phi\right\}\left(1-i\gamma_5\right).
\end{array}\label{eqVI.12}
\end{equation}
The EK wave functions are given by
\begin{equation}\begin{array}{c}
\psi_{EK}^{(+)}(\bm x,t)=
e^{-iEt}\left\{1-\frac{p^2}{8m^2}-\frac{1}{16m^2}\left[({\cal
F}-V)p^2+2\bm\Sigma\cdot\bm p\,({\cal F}-V)\bm\Sigma\cdot\bm
p+p^2({\cal F}-V)\right]
\right.\\\left.+\frac{i}{4}\{\bm\Sigma\cdot\bm p,{\cal
F}\}\frac{1}{mV}-\frac{1}{4m}\bm{\Sigma}\cdot\bm\phi\right\}\left(\begin{array}{c}
\phi^{(+)}(\bm x)
\\ 0\end{array}\right), \\\psi_{EK}^{(-)}(\bm x,t)=
e^{iEt}\left\{1-\frac{p^2}{8m^2}-\frac{1}{16m^2}\left[({\cal
F}-V)p^2+2\bm\Sigma\cdot\bm p\,({\cal F}-V)\bm\Sigma\cdot\bm
p+p^2({\cal F}-V)\right]
\right.\\\left.-\frac{i}{4}\{\bm\Sigma\cdot\bm p,{\cal
F}\}\frac{1}{mV}-\frac{1}{4m}\bm{\Sigma}\cdot\bm\phi\right\}\left(\begin{array}{c}
0 \\ \chi^{(-)}(\bm x) \end{array}\right).
\end{array}\label{eqVIXIII}
\end{equation}
In Eq. (\ref{eqVIXIII}), $E$ is
the module of the Dirac particle energy in the external
gravitational field which Hamiltonian is defined by Eq.
(\ref{eqVI.10}).

Thus, reduction condition (\ref{eq5.1}) is not satisfied for the
EK wave functions. This conclusion remains valid for the EK
transformation performed in Ref. \cite{Ob3} for Dirac particles
interacting with a plane gravitational wave and a constant uniform
magnetic field.

A closed transformation of the EK type was applied in Ref.
\cite{Heidenreich} to Hamiltonian (\ref{eqVI.10}) using the
supersymmetric quantum mechanics methods. The resultant
Hamiltonian coincides with the transformed
one obtained in Refs.
\cite{Ob1,Ob2} in any order of the expansion in powers of $1/m$.

The transformation operator used in Ref. \cite{Heidenreich} is
\begin{equation}
U=\frac{1}{\sqrt2}\left(1+\beta\frac{Q}{(Q^2)^{1/2}}\right)\frac{1}{\sqrt2}(1-i\gamma_5),
~~~Q=\frac12\{\bm\alpha\cdot\bm p,{\cal F}\}+i\gamma_5\beta mV.
\label{eqVI.14}
\end{equation}
Expanding Eq. (\ref{eqVI.14}) in a series up to the terms of the
first order in the potentials $(V-1)$, $({\cal F}-1)$ and their
first spatial derivatives and taking into account only terms up to
$\sim 1/m^2$, one can prove that
formulas (\ref{eqVI.12}) and
(\ref{eqVI.14}) for the transformation operator coincide within
the accepted accuracy. Similarly to the transformation applied in
Ref. \cite{Ob1}, reduction condition (\ref{eq5.1}) is not
satisfied. Hence, the transformation constructed in Ref.
\cite{Heidenreich} is not the FW one.

\subsection{Generalized Melosh transformation}

The Melosh transformation \cite{Me} is often used in the strong
interaction theory. It has been independently proposed by Tsai
\cite{T} to describe interactions of particles with spins 1/2 and
1 with the magnetic field. For free particles, the exponential
operator transforming Dirac Hamiltonian (\ref{eq3.3}) is given by
\cite{T}
\begin{equation}
U_1=\exp{\left(-\frac12\arctan{\frac{\bm\gamma\cdot\bm
p_\bot}{m}}\right)}, ~~~ \bm p_\bot=p_x\bm e_x+p_y\bm e_y.
\label{eqVI.15}
\end{equation}
This transformation operator can be expressed in the equivalent form
\begin{equation}
U_1=\frac{\epsilon+m+\bm\gamma\cdot\bm
p_\bot}{\sqrt{2\epsilon(\epsilon+m)}}, ~~~\epsilon=\sqrt{m^2+\bm
p_\bot^2}. \label{eqVIXVI}
\end{equation}
The transformed Hamiltonian is \cite{Me,T}
\begin{equation}
{\cal H}_1=\beta(\epsilon+\gamma_zp_z). \label{eqVI.17}
\end{equation}

This Hamiltonian is not block-diagonal. To bring it to the
block-diagonal form, one can perform the second transformation.
Since the form of Hamiltonian (\ref{eqVI.17}) is covered by the
condition of exact FW transformation (\ref{eq3.9}), the second
transformation operator is equal to
\begin{equation}
U_2=\frac{E+\epsilon+\gamma_zp_z}{\sqrt{2E(E+\epsilon)}},
~~~E=\sqrt{\epsilon^2+p_z^2}=\sqrt{m^2+\bm p^2}. \label{eqVI.18}
\end{equation}
The resultant transformation operator is given by
\begin{equation}
U_M=U_2U_1. \label{eqVI.19}
\end{equation}
It brings initial Dirac Hamiltonian (\ref{eq3.3}) to the form
\begin{equation}
{\cal H}_M=\beta\sqrt{m^2+\bm p^2}=\beta E. \label{eqVI.20}
\end{equation}
Despite the block-diagonality of
generalized Melosh transformation (\ref{eqVI.19}), it is not
equivalent to the FW one. The connection between the
generalized Melosh transformation and the FW one is given by
\begin{equation}\begin{array}{c}
U_{M\rightarrow
FW}=\frac{\sqrt{(E+\epsilon)(\epsilon+m)}+i\sqrt{(E-\epsilon)(\epsilon-m)}R}
{\sqrt{2\epsilon(E+m)}},\\
U_{FW\rightarrow
M}=\frac{\sqrt{(E+\epsilon)(\epsilon+m)}-i\sqrt{(E-\epsilon)(\epsilon-m)}R}
{\sqrt{2\epsilon(E+m)}},~~~R=\frac{p_x\sigma_y-p_y\sigma_x}{\sqrt{p_x^2+p_y^2}}.
\end{array}\label{eqVI.21}\end{equation}
The wave functions in the generalized Melosh representation are
equal to
\begin{equation}\begin{array}{c}
\psi_{M}^{(+)}(\bm x,t)=U_M\psi_{D}^{(+)}(\bm x,t)=
e^{-iEt}\frac{\sqrt{(E+\epsilon)(\epsilon+m)}-i\sqrt{(E-\epsilon)(\epsilon-m)}R}
{\sqrt{2\epsilon(E+m)}}\left(\begin{array}{c} \phi^{(+)}(\bm x)
\\ 0\end{array}\right),\\
\psi_{M}^{(-)}(\bm x,t)=U_M\psi_{D}^{(-)}(\bm x,t)=
e^{iEt}\frac{\sqrt{(E+\epsilon)(\epsilon+m)}-i\sqrt{(E-\epsilon)(\epsilon-m)}R}
{\sqrt{2\epsilon(E+m)}}\left(\begin{array}{c} 0 \\ \chi^{(-)}(\bm
x) \end{array}\right). \end{array}\label{eqVI.XXII}
\end{equation}
Therefore, these wave functions do not satisfy reduction condition
(\ref{eq5.1}).

\section{Applications of connection between
the Dirac and Foldy-Wouthuysen wave functions}

The Hamiltonian for relativistic particles in the FW
representation contains a square root of operators (see Refs.
\cite{FW,JMP}). Therefore, the Dirac representation is usually
more convenient than the FW one for finding wave eigenfunctions
and eigenvalues of the Hamilton operator. Many exact solutions of
relativistic wave equations have been found just in the Dirac
representation \cite{BG}. Nevertheless, a derivation of equations
of motion is much more difficult in this representation than in
the FW one \cite{JMP,CMcK}.

The use of connection between wave functions in the Dirac and FW
representations defined by Eq. (\ref{eq5.1}) is very important.
One can calculate wave eigenfunctions in the Dirac representation
and then obtain corresponding eigenfunctions in the FW
representation. After that, one can determine expectation values
of needed operators corresponding to certain classical quantities
and derive quantum and semiclassical equations of motion. When the
semiclassical approximation is not admissible, quantum formulae
describing the evolution of the operators can be derived.
Semiclassical evolution of classical quantities corresponding to
these operators can be obtained by averaging the operators in the
solutions. An example of such an evolution is time dependence of
average energy and momentum in a two-level system. Another example
is the spin dynamics in external fields. It is very difficult to
solve these problems in the Dirac representation. It is important
that Eq. (\ref{eq5.1}) is exact because one can solve the above
mentioned problems with any desirable accuracy. The example of
description of spin evolution in the FW representation has been
given in Ref. \cite{PEPANFW}.

In the Dirac representation, the connection between operators and
classical quantities is rather complicated and sometimes not
clear. Explicit expressions for the operators that correspond to
certain classical quantities are known only for free relativistic
particles (see. \cite{FW}). It is clear that the expressions for
these operators in the general case must depend on the parameters
that characterize the external field. The FW transformation if
free of this drawback. Main operators including the operators of
position, momentum, and spin have the same form as in the
nonrelativistic quantum theory. The determination of the FW wave
function allows calculating, e.g., expectation values of operators
of the root-mean-square radius $\sqrt{\langle r^2\rangle}$,
electric and magnetic dipole moments, kinetic energy and so on. In
particular, the operators of the electric and magnetic dipole
moments are equal to $\partial{\cal H}_{FW}/\partial \bm E$ and
$\partial{\cal H}_{FW}/\partial \bm B$, respectively. The use of
the FW representation for this purpose can be effective not only
in the atomic physics but also in the nuclear and particle
physics.

\section{Summary}

The paper formulates and proves the conditions enabling the
transition from the Dirac representation to the FW one. An exact
correlation between wave functions in both representations has
been established. It has been demonstrated that the
block-diagonalization of the Hamiltonian is often insufficient
(see Refs. \cite{Ob1,Ob2,Heidenreich,Ob3,Me}) for its
transformation to the FW representation. Such a transformation
becomes possible, if the reduction condition (\ref{eq5.1}) is
satisfied. The results obtained enable unambiguous transition to
the FW transformation and calculation of matrix elements
and expectation values of the
operators that correspond to the major classical quantities. It is
possible because the exact form of such operators in the FW
representation -- as opposed to the Dirac representation -- can be
established quite easily.

\section*{Acknowledgements}

This work was supported in part by the Belarusian Republican Foundation for
Fundamental Research (grant No. $\Phi $08D-001).

\end{document}